\documentclass[preprint,prb,amsmath]{revtex4}

\usepackage{graphicx}
\usepackage{epsfig}

\begin{document}
\title{
Exact results for the infinite supersymmetric extensions\\
of the infinite square well}
\author{K. Gutierrez, E. Le\'{o}n, M. Belloni}
\email{mabelloni@davidson.edu}
\affiliation{Department of Physics, Davidson College, Davidson, North Carolina
28035 USA}

\author{R. W. Robinett}
\email{rq9@psu.edu}
\affiliation{Department of Physics, The Pennsylvania State
University, University Park, Pennsylvania 16802 USA}

\begin{abstract}

One-dimensional potentials defined by $V^{(S)}(x) =S(S+1) \hbar^2 \pi^2 /[2ma^2\sin^2(\pi x/a)]$ (for integer $S$) arise in the repeated supersymmetrization of the infinite square well, here defined over the region $(0,a)$.  We review the derivation of this hierarchy of potentials and then use the methods of supersymmetric quantum mechanics, as well as more familiar textbook techniques, to derive compact closed-form expressions for the normalized solutions, $\psi_n^{(S)}(x)$,  for all $V^{(S)}(x)$ in terms of well-known special functions in a pedagogically accessible manner.   We also note how the
solutions can be obtained as a special case of a family of shape-invariant potentials, the trigonometric P\"oschl-Teller potentials, which can be used to confirm our results.  We then suggest additional avenues for research questions related to, and pedagogical applications of, these solutions, including the behavior of the corresponding
momentum-space wave functions $\phi_n^{(S)}(p)$ for large $|p|$ and general questions about the supersymmetric hierarchies of potentials which include an infinite barrier.

\end{abstract}

\maketitle

\section{Introduction}
\label{sec:intro}

Generations of physicists have been trained in quantum mechanics by repeated practice using a handful (literally, five or so) of familiar model systems, including the infinite square well (hereafter ISW), harmonic oscillator (HO), hydrogen atom, and rigid rotator. Some of these systems, including many with a high degree of symmetry, are amenable to operator methods, such as the raising and lowering operator approach for the HO. These systems, however, are typically solved using standard techniques involving the solution of the Schr\"odinger equation in position space, enforcing the appropriate boundary conditions to give the quantized energy levels, and often  finding solutions in terms of special functions. Given the very limited number of such tractable examples, it can be a welcome addition to the literature of introductory quantum mechanics to find an entirely new class of potentials which can be approached (and solved completely) using a variety of such methods.

One method of obtaining `new potentials from old' is the use of supersymmetric quantum
mechanics\cite{witten,cooper,cooper_khare_sukhatme_1,cooper_khare_sukhatme_2,russiansusy,sukumar,ejp1,ejp2} (SUSYQM) which can generate new model systems which have the same energy eigenvalue spectrum, save for missing the ground-state energy in the new `supersymmetrized' version of the potential, the so-called partner or superpartner potential. For the hydrogen atom, the supersymmetrization leads to  a hierarchy\cite{cooper_khare_sukhatme_2,sukumar_hierarchy,khare_conference} of potentials all related to each other by the supersymmetrization process.

It has been known for some time (if not generally appreciated) that the most familiar of all model systems, the infinite  square well, has just such a hierarchy\cite{cooper_khare_sukhatme_2,sukumar_hierarchy} of  superpartner potentials (with energy spectra related by supersymmetry) with a very simple form, namely
\begin{equation}
V^{(S)}(x) =\frac{S(S+1) \hbar^2 \pi^2}{2ma^2\sin^2(\pi x/a)}
\qquad
\quad
\mbox{for $0<x<a$}
\, ,
\label{susy_isw}
\end{equation}
where the original potential $V^{ISW}(x)$ (or $S\equiv 0$ case) is defined over the interval $(0,a)$. In this notation
$S=1$ corresponds to the first supersymmetrization of the ISW, $S=2$ the result of supersymmetrizing $V^{(S=1)}(x)$,
and so forth. Beyond a mention of the ground-state wave functions  for this specific class of potentials,\cite{cooper_khare_sukhatme_2,khare_conference}
and their relation to the infinite square well, we have found no detailed discussions of the solutions for this hierarchy, in the  research, mathematical physics, or pedagogical literature. For that reason, an exploration of this system, both in the derivation of the solutions, and their interpretation, is the topic of this work where we present closed-form expressions for all solutions, $\psi_n^{(S)}(x)$, of this problem in terms of known special functions. We emphasize throughout the interplay of the application of formal methods from supersymmetric quantum mechanics, more standard (textbook level) approaches using differential equations, and the use of symbolic manipulation tools as pedagogical methods in approaching this problem.

Mathematically sophisticated readers might recognize (or suspect and then confirm) that the  $V^{(S)}(x)$ given above is in a class of shape-invariant potentials described as
(trigonometric) P\"oschl-Teller (hereafter PT1) potentials which have been discussed in the context of supersymmetric quantum mechanics.\cite{cooper_khare_sukhatme_1, sukhatme} General solutions have been derived for this problem and one can reproduce many of the results presented in this work, obtained here by using `textbook' techniques and more pedagogically accessible methods, by using special values of parameters in the PT1 potentials. We show in the Appendix
how the two approaches can be related, for those interested in connections to the mathematical physics literature, but we focus on
a `bottom up' discussion of the problem  which can then be compared to the `top down' solutions using those results. We also note that several of the most familiar, and physically applicable, quantum-mechanical problems, such as the harmonic oscillator or Coulomb potentials, can be considered special cases of more general problems with solutions given by confluent hypergeometric functions, but for pedagogical reasons, such systems are most often derived in a classroom setting using less advanced mathematical `technologies',  such as employed in this work.

In the next section (\ref{section_SUSY}) we review the methods of supersymmetric quantum mechanics and
then in Sec.~\ref{sec:ISW} we revisit the derivation of the fact that the potentials in Eq.~(\ref{susy_isw}) are the result of repeated supersymmetrization of the ISW, also showing how exact normalized solutions to the general $S$ case can be obtained by iteration. In Sec.~\ref{sec:wavefunctions} we use standard textbook methods based on differential equations to derive compact, closed-form solutions for this hierarchy of potentials, for general $S$, which are written in terms of the Gegenbauer polynomials. Finally, in Sec.~\ref{section_conclusions}, we
briefly discuss avenues for further exploration of this rich system, as well as open questions regarding the hierarchies
of supersymmetric extensions of other familiar one-dimensional systems containing infinite wall potential.

\section{Supersymmetric quantum mechanics}
\label{section_SUSY}

Factorization methods\cite{schrodinger_factorization,infeld_and_hull} have historically proved to be powerful tools in the solution of a wide variety of problems in quantum mechanics and mathematical
physics, especially those with a high degree of symmetry. The connection to
supersymmetry\cite{witten,cooper,cooper_khare_sukhatme_1,cooper_khare_sukhatme_2,russiansusy,sukumar,sukhatme,ejp1,ejp2}   (hereafter SUSY) and the interest in
iso-spectral Hamiltonian systems has provided further motivation for using such approaches in a variety of
one-dimensional model systems.

We begin by assuming a generic one-dimensional potential, $V(x)$, admitting a non-degenerate
ground-state solution, $\psi_{0}(x)$, with energy, $E_0$. If we define a shifted
potential energy function, $V^{(-)}(x) \equiv V(x) -E_0$, we know, by construction, that $\psi_0(x)$
satisfies
\begin{equation}
\hat{H}^{(-)} \psi_0(x) \equiv \left[ -\frac{\hbar^2}{2m} \frac{d^2}{dx^2}
+V^{(-)}(x) \right] \psi_0(x) = 0\;,
\label{groundstate}
\end{equation}
and that $V^{(-)}(x)$ has a zero-energy ($E_{0}^{(-)} = 0$) ground-state.
Since $\psi_0(x)$ is assumed known, we can use Eq.~(\ref{groundstate}) to then write $\hat{H}^{(-)}$ in the form
\begin{equation}
\hat{H}^{(-)} = \frac{\hbar^2}{2m}\left[-\frac{d^2}{dx^2} +
\frac{\psi''_0(x)}{\psi_0(x)}\right]\,.
\end{equation}
If we define the ladder operators
\begin{equation}
\hat{A} \equiv \frac{\hbar}{\sqrt{2m}}
\left(\frac{d}{dx} - \frac{\psi_0'(x)}{\psi_0(x)}\right)
\qquad
\mbox{so that}
\qquad
\hat{A}^{\dagger} \equiv \frac{\hbar}{\sqrt{2m}}
\left(- \frac{d}{dx} - \frac{\psi_0'(x)}{\psi_0(x)}\right)\;,
\label{susy-ladder-operators}
\end{equation}
we then have $\hat{A}^{\dagger}\hat{A} = \hat{H}^{(-)}$ and
$\hat{H}^{(-)}$ is factorizable.

While $\hat{A}^{\dagger}\hat{A} = \hat{H}^{(-)}$
now factorizes the original Hamiltonian (up to an additive constant, $-E_0$),
the related combination, $\hat{A}\hat{A}^{\dagger}$,
can be seen to define an (in principle) entirely new potential,  since
\begin{equation}
\hat{A}\hat{A}^{\dagger} \equiv \hat{H}^{(+)}
= -\frac{\hbar^2}{2m} \frac{d^2}{dx^2} + V^{(+)}(x)\;,
\end{equation}
where
\begin{equation}
V^{(+)}(x)  =  V^{(-)}(x) - \frac{\hbar^2}{m} \frac{d}{dx}
\left[\frac{\psi'_0(x)}{\psi_0(x)}\right]
 =  -V^{(-)}(x) + \frac{\hbar^2}{m}
\left[\frac{\psi_0'(x)}{\psi_0(x)}\right]^2
\, .
\label{susy_partner_potential}
\end{equation}

If $\psi_n^{(-)}(x)$ is
any eigenfunction of $\hat{H}^{(-)}$ with eigenvalue
$E_n^{(-)}$, then
$\hat{A} \psi_n^{(-)}(x)$ is an eigenfunction of $\hat{H}^{(+)}$ with
the same eigenvalue. This is easily seen since
\begin{equation}
\hat{H}^{(+)} \left(\hat{A}\psi_n^{(-)} \right)
=  \hat{A}\hat{A}^{\dagger} \left(\hat{A}\psi_n^{(-)}\right) \nonumber \\
=  \hat{A} \left(\hat{H}^{(-)} \psi_n^{(-)}\right)  \nonumber \\
=  E_n^{(-)} \left(\hat{A} \psi_n^{(-)} \right)\, .
\label{connections}
\end{equation}
Similarly one can show that if $\psi_n^{(+)}(x)$ is an eigenfunction of
$\hat{H}^{(+)}$ with eigenvalue $E_n^{(+)}$, then $\hat{A}^{\dagger}
\psi_n^{(+)}(x)$ is an eigenfunction of
$\hat{H}^{(-)}$ with the same eigenvalue.  Taken together, these relations
can be shown to imply\cite{sukhatme}
\begin{equation}
E_n^{(+)}  =  E_{n+1}^{(-)}
\, ,
\quad
\psi_n^{(+)}(x)  =  \frac{1}{\sqrt{E_{n+1}^{(-)}}} \hat{A} \,\psi_{n+1}^{(-)}(x)
\, ,
\quad
\mbox{and}
\quad
\psi_{n+1}^{(-)}(x) = \frac{1}{\sqrt{E_{n}^{(+)}}}
\hat{A}^{\dagger}\, \psi_{n}^{(+)}(x)
\, .
\label{raising_and_lowering}
\end{equation}
Thus, the two systems defined by $V^{(\pm)}(x)$ have the same energy
spectrum, $E_{n}^{(\pm)}$, except that the zero-energy ground-state of $V^{(-)}(x)$ has no counterpart in $V^{(+)}(x)$. We also note that if the original $\psi_{n}^{(-)}(x)$ are orthogonal and normalized, then so are the $\psi_{n}^{(+)}(x)$, since using Eq.~(\ref{raising_and_lowering}) we have
\begin{eqnarray}
\langle \psi_{n}^{(+)} | \psi_{m}^{(+)}\rangle
& = & \frac{1}{\sqrt{  E_{n+1}^{(-)} E_{m+1}^{(-)} }} \langle \psi_{n+1}^{(-)} | \hat{A}^{\dagger}\, \hat{A} |
\psi_{m+1}^{(-)} \rangle \nonumber \\
& = &
\frac{1}{\sqrt{  E_{n+1}^{(-)} E_{m+1}^{(-)} }}
\langle \psi_{n+1}^{(-)} | \hat{H}^{(-)} |
\psi_{m+1}^{(-)} \rangle \nonumber \\
 & = & \sqrt{\frac{E_{m+1}^{(-)}}{E_{n+1}^{(-)}} }\, \langle \psi_{n+1}^{(-)} | \psi_{m+1}^{(-)} \rangle
= \delta_{n,m}
\, .
\label{normalization_kept}
\end{eqnarray}

As an example, we note that the simplest SUSYQM version of a familiar one-dimensional system is the harmonic oscillator (HO), with potential energy and energy eigenvalues given by
\begin{equation}
V^{\textrm{HO}}(x) = \frac{1}{2}m\omega^2 x^2
\qquad
\mbox{and}
\qquad
E_n = (n+1/2)\hbar \omega
\,,
\label{sho_potential}
\end{equation}
where $n=0,1,2...$.
The energy-eigenstate solutions are well-known to be
\begin{equation}
\psi_{n}(x) = C_n H_{n}(v)\, e^{-v^2/2}
\, ,
\label{sho_solutions}
\end{equation}
where $v \equiv x/\beta$, $\beta \equiv \sqrt{\hbar/m \omega}$, the $H_n(v)$ are the Hermite polynomials, and the
$C_n$ are normalization constants given by $C_n = 1/\sqrt{\beta \sqrt{\pi} 2^n n!}$.

To apply the methods above, we first `zero out' the potential and energy eigenvalues
by subtracting $E_{0} = \hbar \omega/2$ from both to obtain
\begin{equation}
V^{(-)}(x) = \frac{1}{2}m\omega^2 x^2 - \frac{\hbar \omega}{2}
\qquad
\mbox{and}
\qquad
E_n^{(-)} = n\hbar \omega
\, .
\end{equation}
Then since $\psi_{0}(x) \propto e^{-x^2/2\beta^2}$ we find that the superpartner potential is
\begin{equation}
V^{(+)}(x) = - V^{(-)}(x) + \frac{\hbar^2}{m} \left( - \frac{x}{\beta^2}\right)^2
= \left(- \frac{1}{2}m\omega^2 x^2 + \frac{\hbar \omega}{2}\right) +  m \omega^2 x^2
= \frac{1}{2}m\omega^2x^2 +  \frac{\hbar \omega}{2}
\, ,
\end{equation}
with corresponding energies given by
\begin{equation}
E_n^{(+)} = E_{n+1}^{(-)} = (n+1)\hbar \omega = (n+1/2)\hbar \omega + \frac{\hbar \omega}{2}
\, .
\end{equation}
We see that up to the common constant energy term, $\hbar \omega /2$, $V^{(+)}(x)$ and $E_n^{(+)}$ are exactly the same as for the original harmonic oscillator system, so that the supersymmetric partner potentials are in fact identical.
Using the results of Eq.~(\ref{raising_and_lowering}) for the wave functions, we find that
\begin{equation}
\psi^{(+)}_{n}(x) = \frac{1}{\sqrt{E_{n+1}^{(-)}}} \hat{A}\, \psi_{n+1}^{(-)}(x)
=\frac{1}{\sqrt{n+1}} \left(\frac{\beta}{\sqrt{2}}\right)\left(\frac{d}{dx}
+  \frac{x}{\beta^2}\right) \psi_{n+1}^{(-)}(x)
\, ,
\end{equation}
which is equivalent to the standard textbook lowering operator relation,
\begin{equation}
|n\rangle =\frac{1}{\sqrt{n+1}}\,\hat{A} |n+1 \rangle
\qquad
\mbox{where}
\qquad
\hat{A} \equiv\frac{1}{\sqrt{2 m\hbar\omega}} [i\hat{p} +m\omega x]\,.
\end{equation}

\section{Supersymmetric versions of the infinite square well}
\label{sec:ISW}

In contrast to the harmonic oscillator, the supersymmetric partner of the infinite square well (ISW), is non-trivially
different, so we begin by examining the first `SUSY extension' of the ISW, which we label by $S=1$. We define the ISW potential by\cite{de_vincenzo}
\begin{equation}
V^{\textrm{ISW}}(x) \equiv
\begin{cases}
\infty & \mbox{for $x\leq 0$ or $x\geq a$} \\
0      & \mbox{for $0<x<a$}
\end{cases}
\label{ISWpotential}
\end{equation}
and label the energy eigenstates and eigenvalues as
\begin{equation}
\psi_{n}^{\textrm{ISW}}(x) \equiv  \psi_{n}(x) =
\begin{cases}
 0 & \mbox{for $x \leq 0$ or $x \geq a$} \\
\sqrt{\frac{2}{a}} \sin \left[\frac{(n+1)\pi x}{a}\right]      & \mbox{for $0<x<a$}
\end{cases}
\qquad
\mbox{and}
\qquad
E_n = \frac{\hbar^2 \pi^2}{2ma^2}(n+1)^2
\, ,
\label{standard_ISW}
\end{equation}
where $n=0,1,2,...$ so that the label $n=0$ corresponds to the ground-state, to be consistent with the notation in
Sec.~\ref{section_SUSY}. A parameter that will appear often in subsequent expressions is the zero-point energy of the ISW, which we will define as ${\cal E}_0 \equiv E_0=\hbar^2 \pi^2/2ma^2$.
By subtracting this zero-point energy from the potential and energy eigenvalues, we have
\begin{equation}
V^{(-)}(x) =V^{ISW}(x) - {\cal E}_0
\qquad
\mbox{and}
\qquad
E_{n}^{(-)} = {\cal E}_0\left[(n+1)^2-1\right]
\,.
\end{equation}
The ground-state wave function is
$\psi_{0}(x) =\sqrt{2/a}\sin(\pi x/a)$
which gives
\begin{equation}
\left(\frac{\psi_{0}'(x)}{\psi_{0}(x)}\right)^2 =
\frac{\pi^2}{a^2} \frac{\cos^2(\pi x/a)}{\sin^2(\pi x/a)}
\label{ratio}
\, ,
\end{equation}
so that using Eq.~(\ref{susy_partner_potential}) and Eq.~(\ref{ratio}), the partner potential to $V^{(-)}(x)$ is
\begin{equation}
V^{(+)}(x) =   {\cal E}_0 \frac{2}{\sin^2(\pi x/a)}  - {\cal E}_0
\, ,
\end{equation}
with corresponding energies given by
\begin{equation}
E_{n}^{(+)} = E_{n+1}^{(-)} =  {\cal E}_0(n+2)^2 - {\cal E}_0\, .
\end{equation}
Just as in the harmonic oscillator case, since both $V^{(-)}(x)$ and $E_{n}^{(-)}$ have the common factor of
$-{\cal E}_0$, we can rescale the zero of potential and quantized energies to find that the `first supersymmetrization' (or $S=1$ version) of the ISW can be re-defined as
\begin{equation}
V^{(S=1)}(x) =\frac{2{\cal E}_0}{\sin^2(\pi x/a)} =\frac{1(1+1){\cal E}_0}{\sin^2(\pi x/a)}
\quad \,
\mbox{and}
\quad \,
E_{n}^{(S=1)} = {\cal E}_0(n+2)^2
\, .
\end{equation}
This  result has appeared in numerous journal articles,\cite{sukumar,sukhatme}
monographs\cite{cooper,mallow} and even has made its way into textbook problem
sets.\cite{schwabl_textbook,robinett_textbook} Completely independently of the SUSYQM connection to the ISW, discussions of similar potentials
have appeared in collections of quantum mechanics problems\cite{problems} and ultimately can trace its origin back to the
(trigonometric) P\"oschl-Teller potential\cite{poschl_teller} and at least one group\cite{astorga} has explored the SUSY partners of
that case.

Using the results of Eq.~(\ref{raising_and_lowering}) we find the wave functions of the $S=1$ system to be
\begin{eqnarray}
\psi_{n}^{(S=1)}(x) & = & \frac{1}{\sqrt{E_{n+1}^{(-)}}}\, \hat{A}\, \psi_{n+1}^{\textrm{ISW}}(x)\nonumber  \\
& = & \frac{a}{\pi} \frac{1}{\sqrt{(n+2)^2-1}}
\left(\frac{d}{dx} - \frac{\pi}{a}\frac{\cos(\pi x/a)}{\sin(\pi x/a)} \right)\psi_{n+1}^{\textrm{ISW}}(x)
\label{S_1_iteration} \\
& = & \sqrt{\frac{2}{a}}\frac{1}{\sqrt{(n+2)^2-1}}
\left\{(n+2) \cos\left[\frac{(n+2)\pi x}{a}\right] - \frac{\cos( \pi x/a)}{\sin( \pi x/a)}
\sin\left[\frac{(n+2)\pi x}{a}\right]
\right\} \nonumber
\, .
\end{eqnarray}
For comparison to results from the ISW (or $S=0$) case and higher $S$ solutions, we note that for $S=1$ the $n=0,1$ solutions are (up to an arbitrary sign factor) given by
\begin{equation}
\psi_{0}^{(S=1)}(x)  =  2 \sqrt{\frac{2}{3a}} \sin^2 \left(\frac{\pi x}{a}\right)
\qquad
\mbox{and}
\qquad
\psi_{1}^{(S=1)}(x)  =      \frac{4}{\sqrt{a}}\cos\left(\frac{\pi x}{a}\right)   \sin^2 \left(\frac{\pi x}{a}\right)
\label{0_and_1} \, .
\end{equation}

One can now repeat  the supersymmetrization procedure by acting on the $S=1$ solutions
(using $\psi_0^{(S=1)}(x)$ from Eq.~(\ref{0_and_1}) as the new ground-state wave function)
to obtain the supersymmetric partner potential
(and their energy eigenvalues) corresponding to $S=2$. After again  taking into account  identical energy factors (common to both the potential and energies), we find that
\begin{equation}
V^{(S=2)}(x) = {\cal E}_0\frac{6}{\sin^2(\pi x/a)} = {\cal E}_0\frac{2(2+1)}{\sin^2(\pi x/a)}
\quad
\mbox{and}
\quad
E_{n}^{(S=2)} = {\cal E}_0(n+3)^2
\, ,
\end{equation}
with the wave functions given by
\begin{equation}
\psi_{n}^{(S=2)}(x) = \frac{a}{\pi} \frac{1}{\sqrt{(n+3)^2-4}}\left(\frac{d}{dx} - \frac{2\pi }{a}
\frac{\cos(\pi x/a)}{\sin (\pi x/a)}\right) \psi_{n+1}^{(S=1)}(x)
\, .
\label{S_2_iteration}
\,
\end{equation}
We see that this gives $\psi_{0}^{(S=2)}(x) \propto \sin^3(\pi x/a)$ and $\psi_{1}^{(S=2)}(x) \propto \cos(\pi x/a)\, \sin^3(\pi x/a)$ which can be compared to the results in Eq.~(\ref{0_and_1}).

One quickly recognizes the pattern, and by repeatedly applying the supersymmetrization procedures, one can show that the family of potentials generated in this hierarchy of supersymmetric extensions of the infinite square well (hereafter SISW) is given by\cite{cooper_khare_sukhatme_2,sukumar_hierarchy}
\begin{equation}
V^{(S)}(x) =  {\cal E}_0\frac{S(S+1)}{\sin^2(\pi x/a)}
\qquad
\mbox{and}
\qquad
E_{n}^{(S)} = {\cal E}_0(n+S+1)^2
\label{general_susy_potential}
\,.
\end{equation}
We illustrate this hierarchy of potential energy functions, with the corresponding energy spectra, in Fig.~1. We note that for quantized energies given by
$k^2{\cal E}_0$  there are $k$ different $V^{(S)}(x)$ potentials which will have that value as a possible state. All of the energy
eigenvalues represented by dashed lines in Fig.~1 which are above the minimum value of a given $V^{(S)}(x)$
 (namely $V_{\mathrm{min}}^{(S)} = V^{(S)}(x=a/2) = S(S+1){\cal E}_0$) correspond to allowed states of that system.

The operator connection which generalizes the results from Eqns.~(\ref{S_1_iteration}) and (\ref{S_2_iteration})
to connect the  $S$ and $S+1$ states is then
\begin{eqnarray}
\psi_{n}^{(S+1)}(x)
& = & \left(\frac{a}{\pi}\right)
\frac{1}{\sqrt{(n+S+2)^2 - (S+1)^2}}
\left[\frac{d}{dx} - \frac{(S+1)\pi}{a}\frac{\cos(\pi x/a)}{\sin(\pi x/a)}\right] \psi_{n+1}^{(S)}(x) \nonumber \\
& \equiv & \hat{B}^{(S)}\, \psi_{n+1}^{(S)}(x)
\, ,
\label{general_iteration}
\end{eqnarray}
which defines the general operator $\hat{B}^{(S)}$, which is  analogous to the SUSYQM operator $\hat{A}$,  but
made dimensionless.

This approach can then (in principle) be used to obtain the energy eigenstates of any $(n,S)$ combination by iteratively using Eq.~(\ref{general_iteration})
as often as necessary to generate the desired state. For example, if one wants the $(n=5,S=7)$ state, one can repeatedly act on the $(n=12,S=0$) state
(i.e., the $n=12$ ISW wave function) using the appropriate $\hat{B}^{S}$ operators, or more generally
\begin{equation}
\psi_{n}^{(S)}(x) = \prod_{p=1}^{p=S} \hat{B}^{(p)}\, \left[\psi_{n+S}^{\textrm{ISW}}(x)\right]\, .
\label{iteration}
\end{equation}
One can, of course, implement  this algorithm in symbolic manipulation programs to extract any desired solution very efficiently.

Because the supersymmetrization procedure respects the normalization
of the wave functions, as shown  in Eq.~(\ref{normalization_kept}), we know that the  $\psi_{n}^{(S)}(x)$ solutions will be
appropriately normalized since the original $\psi_{n}^{(S=0)}(x) = \psi_{n}^{\textrm{ISW}}(x)$ were. Using results obtained
from this approach, we illustrate the lowest-lying
quantum states ($n=0,1,2)$ for the first three values of $S$ (including the $S=0$ or ISW case) in Fig.~2 which exhibit the expected nodal structure.
For larger $S$ values, the probability density for low-$n$ states is preferentially located
the center of the well, and away from the walls at $x=0,a$, in contrast to the more `flat' distribution for the ISW.
To visualize this  limiting case, we plot some of the
$\psi_n^{(S)}(x)$  in Fig.~3, where for fixed $n=5$ we show $|\psi_n^{(S)}(x)|^2$ for two values of $S$ ($S = 0,10$). This more clearly illustrates the peaking of the quantum probability density near the classical turning points, namely where $E_n^{(S)} = V^{(S)}(x)$,
in the non-trivial (non-ISW) $S>0$ cases.

The result for the hierarchy of supersymmetric ISW (SISW) potentials in Eq.~(\ref{general_susy_potential}) was evidently first noted by Sukumar,\cite{sukumar_hierarchy} and has also been discussed by others\cite{cooper_khare_sukhatme_2,khare_conference} who,  in addition, showed that the ground-state wave functions (in our notation) are proportional to $\psi_{0}^{(S)}(x) \propto \sin^{(S+1)}(\pi x/a)$. The ground-state wave functions are automatically generated by the repeated supersymmetrizations above and confirm this result, and we then easily find
the completely normalized results for $\psi_{0}^{(S)}(x)$. One can also easily see that the first-excited states are proportional to $\cos(\pi x/a) \, \sin^{(S+1)}(\pi x/a)$ (again consistent with earlier results) and obtain the corresponding normalizations for them as well.  In this way we find the universal  result for the ground-state and first-excited state  for the general $S$ case is
\begin{eqnarray}
\psi_{0}^{(S)}(x) & = & \frac{1}{\sqrt{a}}\left[\frac{\sqrt{\pi}\,\Gamma(S+2)}{\Gamma(S+3/2)}\right]^{1/2}
\sin^{S+1}(y) \label{general_0}\\
\psi_{1}^{(S)}(x) & = & \frac{1}{\sqrt{a}}\left[\frac{2\sqrt{\pi}\,\Gamma(S+3)}{\Gamma(S+3/2)}\right]^{1/2}
\cos(y) \, \sin^{S+1}(y)\, , \label{general_1}
\end{eqnarray}
where we will henceforward write  $ y \equiv\pi x/a$ for notational simplicity. These results can be confirmed by direct substitution into the
 Schr\"odinger equation
for $V^{(S)}(x)$ and $E_n^{(S)}$ from Eq.~(\ref{general_susy_potential}), providing an example of the pedagogical use of many aspects of this
rich problem.
It is also easy to show that the $\psi_{0}^{(S+1)}(x)$ and $\psi_{1}^{(S)}(x)$ satisfy the operator relation $\psi_{0}^{(S+1)}(x) = \hat{B}^{(S)}
\, \psi_{1}^{(S)}(x)$
in Eq.~(\ref{general_iteration}) for general $S$.

We note that for $x\approx 0$ (i.e. near the infinite wall), the potential for the general $S$ case reduces to
\begin{equation}
V_{(S)}(x) \sim \frac{S(S+1) \hbar^2}{2mx^2}
\, ,
\label{centrifugal}
\end{equation}
which is clearly similar in form to the standard `centrifugal barrier' term arising from angular momentum considerations in 3D problems involving central potentials, namely $V_C(r) = l(l+1)\hbar^2/2mr^2$, here with the parameter $S$ playing the role of the angular momentum quantum number $l$: a similar
barrier term also arises near the other  infinite wall at $x=a$.

To understand this behavior, we observe that for such an initial potential the wave function near a wall (say at $x=0$) must have the form $\psi_{0}^{(S=0)}(x) = a_1x + {\cal O}(x^2)$ so that the $S=1$ potential will necessarily contain a term of the form
\begin{equation}
V^{(S=1)}(x \! \sim \! 0) =
\frac{\hbar^2}{m}     \left( \frac{\psi_{0}'(x)}{\psi_{0}(x)} \right)^2
\sim \frac{\hbar^2}{m} \left(\frac{a_1}{a_1 x}\right)^2
\propto \frac{2 \hbar^2}{2mx^2}
\, ,
\end{equation}
giving an $S(S+1) = 1(1+1) = 2$ `centrifugal barrier' term in the first supersymmetric potential near $x=0$. The $S=1$ wave functions must  then satisfy the appropriate boundary conditions at $x \sim 0$ (just as would 3D radial wave functions in central potentials), namely $\psi_{0}^{(S=1)}(x) = a_2x^2 + {\cal O}(x^3)$ and using this dependence when one performs the second supersymmetrization, one finds $V^{(S=2)}(x)  = 6\hbar^2 /2mx^2$
consistent with $S(S+1) = 2(2+1)=6$. Once again, one can proceed by induction to derive the form in Eq.~(\ref{centrifugal}) and the fact that
$\psi_{n}^{(S)}(x\! \sim \! 0) \propto x^{S+1}$. This behavior is clearly illustrated  in Fig.~2 (lower-right frame) where $\psi_0^{(S)}(x)$ approaches
 the boundaries at $x=0$ increasingly smoothly as $S$ increases, as the wave function `tunnels' into the `angular-momentum-like' barriers near the walls.

\section{SISW wave functions}
\label{sec:wavefunctions}

To explore the structure of the solutions in the combined $(n,S)$ space, we first use the iterative procedures outlined above to collect the $5$ lowest lying $S=1$ solutions. Motivated by the forms in Eq.~(\ref{0_and_1}), we use symbolic manipulation software to expand and factor the resulting trigonometric functions in specific ways to obtain the following
\begin{eqnarray}
\psi_{0}^{(S=1)}(x) & = &   2\sqrt{\frac{2}{3a}}  \sin^2(y)   \label{S_1_0}\\
\psi_{1}^{(S=1)}(x) & = &     \frac{4}{\sqrt{a}} \left[\cos(y)\right] \sin^2(y)   \\
\psi_{2}^{(S=1)}(x) & = &      4\sqrt{\frac{2}{15a}} \left[ -1 + 6\cos^2(y)\right]\sin^2(y) \\
\psi_{3}^{(S=1)}(x) & = &        \frac{4}{\sqrt{3a}}\left[  -3\cos(y) + 8\cos^3(y)\right] \sin^2(y)\\
\psi_{4}^{(S=1)}(x) & = &       2\sqrt{\frac{2}{35a}} \left[3 - 48\cos^2(y) + 80 \cos^4(y) \right]\sin^2(y) \, ,
\label{S_1_5}
\end{eqnarray}
where we again use the notation $y\equiv \pi x/a$.  We have done this for higher values of $S$ and find quite generally that all of the solutions for a given value of $S$ have a common factor of $\sin^{(S+1)}(y)$ and that the remaining part of the wave function is a polynomial in $\cos( y)$ of order $n$. This uniform
pattern for the $S \geq 1$ states seems, at first, to be rather different than the standard ISW results in Eq.~(\ref{standard_ISW}) which corresponds to $S=0$, at least until we realize that repeated use of trigonometric identities can be applied to the $\psi^{\textrm{ISW}}_n(x)= \psi_n^{(S=0)}(x)$  to obtain the expressions
\begin{eqnarray}
\psi_{0}^{(S=0)}(x) & = &      \sqrt{\frac{2}{a}}                     \sin(y)   \label{S_0_0} \\
\psi_{1}^{(S=0)}(x) & = &     2\sqrt{\frac{2}{a}}    \left[\cos(y)\right] \sin(y)   \\
\psi_{2}^{(S=0)}(x) & = &       \sqrt{\frac{2}{a}}     \left[ -1 + 4\cos^2(y)\right]\sin(y) \\
\psi_{3}^{(S=0)}(x) & = &     2 \sqrt{\frac{2}{a}}   \left[  -2\cos(y) + 4\cos^3(y)\right] \sin(y)\\
\psi_{4}^{(S=0)}(x) & = &        \sqrt{\frac{2}{a}}    \left[1 - 12\cos^2(y) + 16\cos^4(y) \right]\sin(y)\, ,
\label{S_0_5}
\end{eqnarray}
which are indeed of the same general form.

Building on the similarity between these results and the HO case, we argue that the $\sin^{(S+1)}(y)$ terms here play a role akin to the $e^{-x^2/2\beta^2}$ factors in the HO case, being responsible for `enforcing the boundary conditions.' In the SISW case,
the $\sin^{(S+1)}(y)$ components enforce the boundary conditions at the $x=0,a$ infinite walls, while for the oscillator solutions the Gaussian factors guarantee the smooth vanishing of the wave function  at $x = \pm \infty$.

Motivated by this similarity, we attempt to factor out the $\sin^{(S+1)}(y)$ dependence by writing
\begin{equation}
\psi_{n}^{(S)}(x) =  G_{n}^{(S)}(y)\, \sin^{S+1}(y)
\, ,
\end{equation}
and substituting it into the (dimensionless) Schr\"odinger equation for $V^{(S)}(x)$, namely
\begin{equation}
\frac{d^2\psi_{n}^{(S)}(y)}{dy^2} - \frac{S(S+1)}{\sin^2(y)} \psi_{n}^{(S)}(y) + (n+S+1)^2 \psi_{n}^{(S)}(y) =0
\, ,
\end{equation}
thereby obtaining a differential equation for the $G_{n}(y)$ components given by
\begin{equation}
\sin(y) \frac{d^2 G_{n}^{(S)}(y)}{dy^2} + 2(S+1)\cos(y) \frac{d G_{n}^{(S)}(y)}{dy}
+ \left[(n+S+1)^2 - (S+1)^2\right]\sin(y)G_{n}^{(S)}(y) = 0
\, .
\label{h_equation}
\end{equation}

Using our experience with the form of the solutions for general $(n,S)$, from Eqs.~(\ref{S_1_0}) - (\ref{S_1_5}) and (\ref{S_0_0}) - (\ref{S_0_5}) and beyond, we assume that $G_n^{(S)}(y)$ can be expanded in a
(presumably finite) series in powers of $\cos(y)$, by writing
\begin{equation}
G_{n}^{(S)}[\cos(y)] = \sum_{k=0}^{\infty} a_{k,n} \cos^k(y)
\,.
\end{equation}
Substituting  this into Eq.~(\ref{h_equation}) we find
\begin{equation}
\sum_{k} k(k-1)a_{k,n}\cos^{k-2}(y) = \sum_{k} a_{k,n}[(k+S+1)^2 - (n+S+1)^2]\cos^k(y)
\, ,
\end{equation}
and upon relabeling and comparing similar powers of $\cos(y)$ we find the recursion relation amongst the expansion
coefficients
\begin{equation}
a_{k+2,n} = a_{k,n} \left[\frac{(k+S+1)^2 - (n+S+1)^2}{(k+1)(k+2)}\right]
\, .
\label{recursion_relation}
\end{equation}
This expression confirms that for a given $n$, the series in $\cos(y)$ will indeed  terminate with a highest power of $k=n$. It also connects every {\bf other} term in the expansion, implying that starting with arbitrary $a_{0,n},a_{1,n}$, separate even and odd series will be generated.

This is the identical logic used to conclude that the series expansion for the harmonic oscillator (HO) problem must reduce to a finite polynomial, since otherwise the infinite series would yield the incorrect behavior as $x \rightarrow \pm \infty$. In HO case, the recursion relation of the coefficients one obtains also connects every other term in the expansion, also giving the expected even and odd parity solutions.

The generality of the results obtained so far, namely that the solutions can be constructed from simple factors which
encode the behavior of the solutions at the boundaries, along with polynomials (here in the variable $\cos(y)$) which describe the
dynamical behavior inside the well (the `wiggliness' if you will) suggested to us that these expressions might be able to be mapped onto existing forms in the mathematical literature. Given that the equation for the polynomials yields solutions involving the variable $w = \cos(y)$, we rewrite the differential equation for $G_{n}^{(S)}(y) = F_{n}(w)$ and obtain
\begin{eqnarray}
\sin^2(y) F_{n}''(w) - \cos(y) F'_{n}(w)(2S+3) + n(n+2S+2)F_{n}(w) & =& 0\\
\mbox{or} \qquad \qquad \qquad \qquad &  & \nonumber \\
(1-w^2) F_{n}''(w) - wF'_{n}(w)(2S+3) + n(n+2S+2)F_{n}(w) & = & 0
\, .
\end{eqnarray}
This final form is indeed known in the mathematical physics literature\cite{handbook} as being the equation for the Gegenbauer polynomials, sometimes written in the form
\begin{equation}
(1-z^2) F''(z) - zF'(z)(2\alpha +1) +n(n+2\alpha)F(z) = 0
\, ,
\end{equation}
with solutions expressed in the notation $F(z) = C^{\alpha}_{n}(z)$, where we associate $\alpha = S+1$.

The Gegenbauer functions are polynomials of order $n$, defined over the interval $z \in (-1,+1)$, and mutually orthogonal under the weight $(1-z^2)^{\alpha -2}$. They have $n$ nodes over the allowed range and as $x$ varies from $0$ to $a$ in our physical problem, the argument of  $C_n^{\alpha}[\cos(\pi x/a)]$ varies in the defined range of $(-1,+1)$. The appearance of such orthogonal polynomials should be very familiar from the 1D harmonic oscillator, where the Hermite polynomials, $H_n(z)$, appear with the weight being $e^{-z^2}$ defined over the interval $(-\infty,+\infty)$, and very similar results from the 3D Coulomb problem.

Integrals over the products of the $C_{n}^{\alpha}(z)$ times the appropriate weight functions
\cite{handbook} are exactly the type of  results needed to determine the  normalization of the solutions. Specifically, we find using such results that we can write the general $(n,S)$ solution for the SISW hierarchy of potentials in the form
\begin{equation}
\psi_{n}^{(S)}(x) = \frac{1}{\sqrt{a}}\,
\left[\frac{2^{2S+1} \Gamma(n+1)\Gamma(S+1)^2(n+S+1)}{\Gamma(n+2S+2)}\right]^{1/2}
\, C_{n}^{(S+1)}\left[\cos(y)\right] \sin(y)^{S+1}
\, ,
\label{gegenbauer}
\end{equation}
where again $y \equiv \pi x/a$. Using the standard results for the lowest lying Gegenbauer polynomials, namely
$C_{0}^{\alpha}(y) = 1$ and $C_{1}^{\alpha}(y) = 2\alpha y$,
we can also reproduce our earlier `experimentally derived' results for $\psi_{0,1}^{(S)}(x)$
in Eqns.~(\ref{general_0}) and (\ref{general_1}) for all $S$, including the normalization factors.
In addition, for $S=0$, the results of Eq.~(\ref{gegenbauer}) reduce to a new form of the ISW wave functions
\begin{equation}
\psi_{n}^{(S=0)}(x)=\psi^{\textrm{ISW}}_n(x)=\sqrt{\frac{2}{a}}\;{\cal{U}}_{n-1}[\cos(y)]\;\sin(y)\;,
\label{chebyshev}
\end{equation}
where ${\cal{U}}_n$ are the Chebyshev polynomials of the second kind.

The expression in Eq.~(\ref{gegenbauer}) is one we have not found explicitly in the pedagogical literature,  but we do note that the solutions can be obtained as a special case of more general P\"oschl-Teller potentials and we outline (in some detail) the
connections to those results in an Appendix, for those interested in a more advanced `view' of this problem.
Given the simplicity of this form in Eq.~(\ref{gegenbauer}), we suggest that this system may find a useful place in the teaching of quantum mechanics, especially given the array of additional questions (see Sec.~\ref{section_conclusions} for examples) one can then pursue in analyzing its structure, and as a source of many new examples.

\section{Conclusions and future directions}
\label{section_conclusions}

In this work, we have focused on obtaining the solutions to a novel set of quantum mechanical problems encoded in the hierarchy of supersymmetric extensions of the most familiar of all textbook models, the infinite square well. Using the methods of supersymmetric quantum mechanics, and the mathematical tools outlined in almost every undergraduate quantum textbook, we have been able to present elegant, compact, closed-form solutions to a new class of quantum-mechanical potentials, dramatically extending earlier discussions\cite{cooper_khare_sukhatme_2,sukumar_hierarchy,khare_conference} of this system.
While deriving these  results, we have emphasized  the interplay between various solutions methods in quantum mechanics when approaching new problems, in the same way that any student might when facing `familiar' problems for the first time, so in that sense our work is very pedagogical.

This model system is  now ripe for further study, with many additional areas of research to explore or pedagogical application to use in the classroom.
For example, with the ability to now easily calculate many physical quantities of interest, we have been able to find closed form expressions for the expectation values of the potential and kinetic energies in a general $(n,S)$ state, namely
\begin{eqnarray}
\langle \psi_{n}^{(S)} | V^{(S)}(x) | \psi_{n}^{(S)} \rangle & = &{\cal E}_{0} \left\{\frac{2S(S+1)(n+S+1)}{(2S+1)} \right\}
\label{pot_e}\\
\frac{1}{2m} \langle\psi_{n}^{(S)} |\hat{p}^2| \psi_{n}^{(S)}   \rangle = \langle \psi_{n}^{(S)} |\hat{T}| \psi_{n}^{(S)} \rangle
& = & {\cal E}_{0} \left\{\frac{[(2S+1)n + (S+1)](n+S+1)}{(2S+1)}\right\}
\, ,
\label{ke_e}
\end{eqnarray}
where we find that
\begin{equation}
\langle \psi_{n}^{(S)} | V^{(S)}(x) | \psi_{n}^{(S)} \rangle
+
\langle \psi_{n}^{(S)} |\hat{T}| \psi_{n}^{(S)} \rangle
={\cal E}_{0} (n+S+1)^2 =  E_{n}^{(S)}
\, .
\end{equation}
We have also confirmed (by explicit calculation) that the virial theorem holds, namely that
\begin{equation}
\langle\psi_{n}^{(S) } |\hat{T} |  \psi_{n}^{(S)}  \rangle = \frac{1}{2} \langle \psi_{n}^{(S)} \left| x \frac{dV^{(S)}(x)}{dx} \right|
\psi_{n}^{(S)} \rangle
\, ,
\end{equation}
 for $S \geq 1$ where the potential energy function, $V^{(S)}(x)$, is better behaved than $V^{(S=0)}(x)=V^{\textrm{ISW}}(x)$.
Problems such as these (and many more which suggest themselves) can be used as new classroom examples or homework assignments
(not appearing in standard textbooks) and therefore can certainly be incorporated into an advanced undergraduate class, especially one where computer math tools are encouraged as an educational tool. Other examples include exploration of the Wigner function for this class of potentials, extending existing results for the ISW \cite{wigner}, or time-dependent phenomena such as
wave packet revivals \cite{revivals} where the quadratic energy eigenvalues in Eq.~(\ref{general_susy_potential}) guarantee that exact revivals will
be supported in each $V^{(S)}(x)$ potential.

The simple form of the $\psi_n^{(S)}(x)$ in terms of known special functions suggests that the momentum-space wave functions might also be
written in equally compact and elegant ways. For example, for the 3D Coulomb problem (hydrogen atom) the momentum-space solutions were
deftly derived in the very early days of quantum mechanics\cite{podolsky_and_pauling} in terms of known special functions, in fact Gegenbauer polynomials. We have already started to explore the general $\phi_n^{(S)}(p)$ solutions and have confirmed that they exhibit large $|p|$ behavior
given by $|\phi_n^{(S)}(p)| \sim p^{-(2+S)}$, consistent with theorems\cite{star_wars} connecting  the discontinuities of $\psi(x)$
(here encoded in the increasingly smooth $x^{S+1}$ behavior of the wave functions at the walls)
very directly to the large momentum limit of $\phi(p)$. There are likely many closed-form results waiting to be uncovered in the continued mathematical physics
analysis of both the position-space and momentum-space versions of this problem.

One of the most striking results of the SISW hierarchy is the simple form of the general $V^{(S)}(x)$ potentials, and especially their explicit dependence on the $S(S+1)$ factor.  While we expect the general form in Eq.~(\ref{centrifugal}) near any infinite wall in a SUSY hierarchy, the fact
 that the $S(S+1)$ factor appears as a pre-factor to a relatively simple functional form is perhaps surprising.
We note that two earlier works have already explored the behavior of `half-potential' problems, ones defined by
\begin{equation}
\tilde{V}(x) \equiv
\begin{cases}
\, \infty & \mbox{for $x<0$} \\
V(x)      & \mbox{for $0<x$}
\end{cases}
\label{half_cases}
\,,
\end{equation}
for cases where $V(x)$ has a very high degree of symmetry in the complete 1D case, namely for the
`half-oscillator'   \cite{half_sho} and the 1D Coulomb problem.\cite{yepez} The authors of those studies have  considered (in passing) the $S=1$ supersymmetric extensions of the $S=0$ original potentials for each case and have found
\begin{eqnarray}
V^{(S=0)}(x) = \frac{1}{2}m\omega^2 x^2
& \quad \Longrightarrow \quad &
V^{(S=1)}(x)  =  \frac{1}{2} m\omega^2 x^2 + \frac{2 \hbar^2}{2mx^2} \, ,\\
V^{(S=0)}(x) =  -\frac{Ke^2}{x}
& \quad \Longrightarrow \quad &
V^{(S=1)}(x)  =   -\frac{Ke^2}{x} + \frac{2 \hbar^2}{2mx^2}\, ,
\end{eqnarray}
where the $S=1$ results have not just an approximate $S(S+1)\hbar^2/2mx^2$ behavior near the infinite wall boundary (as suggested by
Eq.~(\ref{centrifugal})), but an exact `centrifugal' term for all $x>0$. We have extended those results and find that repeated
symmetrizations of these two systems give
\begin{eqnarray}
V^{(S=0)}(x) = \frac{1}{2}m\omega^2 x^2
& \quad \Longrightarrow \quad &
V^{(S)}(x)  =  \frac{1}{2} m\omega^2 x^2 + \frac{S(S+1) \hbar^2}{2mx^2} \, ,\\
V^{(S=0)}(x) =  -\frac{Ke^2}{x}
& \quad \Longrightarrow \quad &
V^{(S)}(x)  =   -\frac{Ke^2}{x} + \frac{S(S+1) \hbar^2}{2mx^2}\,
\end{eqnarray}
for the general $S$ case.
We have also found closed-form expressions for the general solutions of these systems, using results from the related fully three-dimensional
versions of the harmonic oscillator and Coulomb problem, where similar potentials occur in the corresponding radial equation. In these cases, the role of the $S$ parameter is indeed closely related to the angular momentum quantum number ($l$) of the 3D problem.

We observe that these two systems are
more examples of shape-invariant potentials, so perhaps the simple forms found in these three cases (the SIWS and the two cases here) are connected to that property. One should note, however, that for the supersymmetric hierarchies of other familiar potentials on the `half-line,' for example the quantum bouncer defined by using $V(x) = Fx$ in Eq.(~\ref{half_cases}), the hierarchy of supersymmetric potentials cannot be written in the simple  form
$V^{(S)}(x) = S(S+1) G(x) + V^{(0)}(x)$, although the form of $V^{(S)}(x)$ near the wall is indeed always of the form in
Eq.~(\ref{centrifugal}). One can, however,  still use the `iterative' approach in Eq.~(\ref{iteration}) to obtain any 
$\psi_{n}^{(S)}(x)$ using as the `seed' ($S=0$) problem the quantum bouncer with normalized solutions given in terms of Airy functions by
$\psi_{n}^{(S=0)}(x) = Ai(x/\rho - \zeta_{n})/Ai'(-\zeta_{n})$ (where the $-\zeta_{n}$ are the zeros of the Airy function).

\appendix*
\section{The SISW as a special case of the (trigonometric)  P\"oschl-Teller shape-invariant potentials}
\label{poschl}

As alluded to above, one can show that the infinite hierarchy of potentials arising from repeated supersymmetrization of the infinite square well potential described by Eq.~(\ref{susy_isw})  (and other results involving the energies and wave functions, as shown in Eqns.~(\ref{general_susy_potential}), (\ref{general_0}), and (\ref{gegenbauer})) can be obtained as a special case of a more general mathematical result involving a family of shape-invariant potentials. For those interested in the details of this connection, we present the necessary background in this Appendix.

Using results from a review of supersymmetric quantum mechanics, \cite{cooper_khare_sukhatme_1} we can write the potential for the so-called (trigonometric) P\"oschl-Teller (hereafter PT1) family of potentials in the form
\begin{equation}
V_{PT1}(x;A,b) \equiv \left( \frac{\hbar^2}{2ma^2} \right)
\left[\frac{A(A-\alpha)}{\cos(\alpha y)^2} + \frac{B(B-\alpha)}{\sin^2(\alpha y)} \right]
\label{PT_potential}
\end{equation}
where $y \equiv x/a$ (as above, but here with the restriction $0\leq y \leq \pi/2\alpha$)  which is dimensionless and one also assumes $A,B,\alpha > 0$. The corresponding energy eigenvalues for this potential are known to be
\begin{equation}
E_n^{PT1} = \left( \frac{\hbar^2}{2ma^2} \right) [A+B+ 2n\alpha]^2
\label{PT_energies}
\, .
\end{equation}
We have modified the standard results~\cite{cooper_khare_sukhatme_1} to reflect the notation used in the current work (including expressing quantities such as the position variable and potential energy and eigenvalues initially in terms of their dimensionful values) as well as adding the same overall constant energy term to both the potential and energy eigenvalues, both  to simplify the result and to make contact with our expressions.

If we define $\sigma \equiv A/\alpha$ and $\tau \equiv B/\alpha$, the ground state of this system is known to be
\begin{equation}
\psi_{0}^{PT1}(x;A,B) \propto [\cos(\alpha y)]^{\sigma} \, [\sin(\alpha y)]^{\tau} \, .
\end{equation}
A second change of variables to $w = 1-2\sin^2(\alpha y)$ then yields a general solution of the form
\begin{equation}
\psi_n^{PT1}(x;A,B) = N_{n}^{(\tau,\sigma)}\,  (1-w)^{\tau/2} \,  (1+w)^{\sigma/2} \, P_{n}^{(\tau-1/2, \sigma -1/2)} (w)
\end{equation}
where the $P_{n}^{(\mu,\nu)}(z)$ are the Jacobi polynomials (themselves special cases of hypergeometric functions) which are orthogonal over the interval $z \in [-1,+1]$ under the weight function $(1-z)^{\mu}\,(1+z)^{\nu}$, specifically
\begin{equation}
\int_{-1}^{+1} (1-z)^{\mu}(1+z)^{\nu}
 P_{n}^{(\mu,\nu)}(z)
 P_{m}^{(\mu,\nu)}(z)
 dz
= H_n\, \delta_{n,m}
\label{jacobi_ortho}
\, .
\end{equation}
Mathematical handbook\cite{handbook} results for $H_n$ can be used to show that the general normalization factor
for $\psi_{n}^{PT1}(x;A,B)$  is given by
\begin{equation}
N_{n}^{(\tau,\sigma)} = \left[
\frac{2\alpha}{a}
\,
\left(\frac{2n+\sigma+\tau}{2^{\sigma+\tau}}\right)
\,
\left(\frac{\Gamma(n+1)\, \Gamma(n+\sigma + \tau)}{\Gamma(n+ \sigma + 1/2) \, \Gamma(n+\tau+1/2)}\right)
\right]^{1/2}
\label{jacobi_normalization}
\,.
\end{equation}

The SISW can now be seen to be a special case of this system if one uses the particular parameters $\alpha = \pi/2$ and $A=B=(S+1)\pi/2$, since in that
case the potential in Eq.~(\ref{PT_potential}) becomes
\begin{eqnarray}
V_{PT1}(x) & = &  \left( \frac{\hbar^2}{2ma^2} \right)
\left[
(S+1)\frac{\pi}{2}\left[(S+1)\frac{\pi}{2} - \frac{\pi}{2}\right]\left\{\frac{1}{\cos^2(\pi y/2)} + \frac{1}{\sin^2(\pi y/2)}
\right\}
\right] \nonumber \\
& = &
\left(\frac{\hbar^2 \pi^2}{2ma^2} \right)\frac{S(S+1)}{[2\cos(\pi y/2) \sin(\pi y/2)]^2}
=
\frac{\hbar^2\pi^2}{2ma^2} \frac{S(S+1)}{\sin^2(\pi y)}
\end{eqnarray}
reducing to Eq.~(\ref{susy_isw}), while the energies from Eq.~(\ref{PT_energies}) are given by
$E_n^{PT1} = (\hbar^2\pi^2/2ma^2) (S+1+n)^2$
in agreement with Eq.~(\ref{general_susy_potential}). In this limit we also have $\sigma = \tau = S+1$ and the ground state wave function  is now
\begin{equation}
\psi_{0}(x) \propto
\left[ \cos(\pi y/2)\right]^{(S+1)/2}
\,
\left[ \sin(\pi y/2)\right]^{(S+1)/2}
\propto \sin(\pi y)^{S+1}
\end{equation}
reproducing the form in Eq.~(\ref{general_0}), while the variable change above reduces to $w= 1- \sin^2(\pi y/2) \equiv \cos(\pi y)$ also consistent with our previous notation. The general solution in this case then becomes
\begin{equation}
\psi_n^{PT1}(x) = N_{n}^{(S+1,S+1)} (1-w)^{(S+1)/2} \, (1+w)^{(S+1)/2} \, P_{n}^{(S+1/2,S+1/2)}(w) \, ,
\end{equation}
and we can also use the fact\cite{handbook} that the Gegenbauer polynomials are special cases of the Jacobi functions, namely
\begin{equation}
C_{n}^{(\nu)}(z) = \left[
\frac{\Gamma(\nu+1/2)\, \Gamma(2\nu+n)}{\Gamma(2\nu) \, \Gamma(\nu + n+1/2)}
\right]
\, P_{n}^{(\nu-1/2,\nu-1/2)}(z)
\, .
\end{equation}
Using these results, along with  $(1-w)^{(S+1)/2} \, (1+w)^{(S+1)/2} = (1-w^2)^{(S+1)/2} = \sin(\pi y)^{S+1}$ and the normalizations in
Eq.~(\ref{jacobi_normalization}), and one gamma function identity, we recover the form derived in Eq.~(\ref{gegenbauer}), including the complete normalizations. (We thank one of the referees for the suggestion that we make this connection to earlier general results explicit in our presentation).

\section{Acknowledgements}

This work was supported in part by a Davidson College Faculty Study and Research Grant (MB and EL), the Davidson College RISE Program (KG), and the Michael D. Jenks Summer Research Fund (MB and KG).

\newpage

\begin{figure}[h!]
\epsfig{file=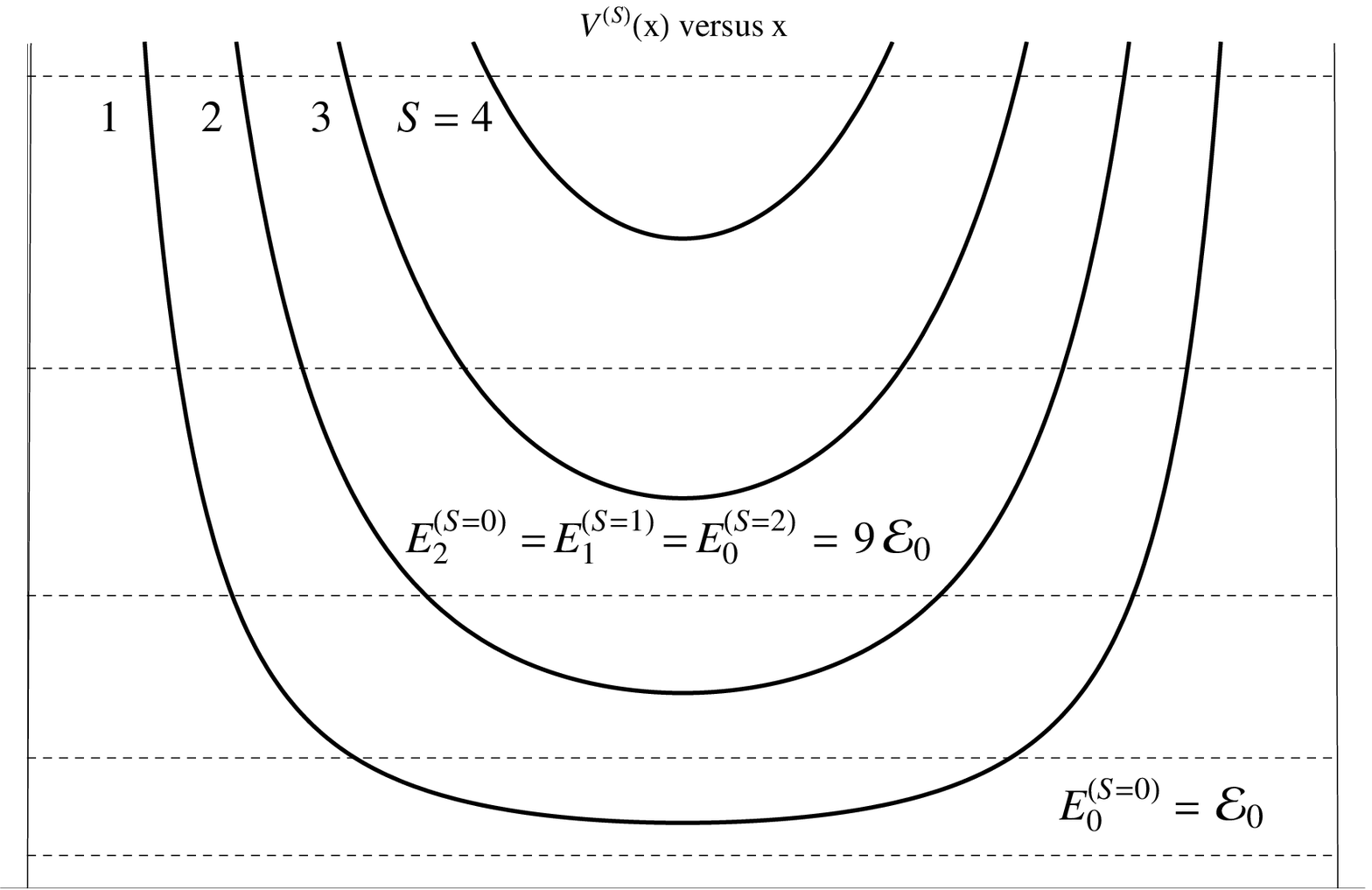,width=0.95\linewidth,angle=0}
\caption{Superpartner potentials, $V^{(S)}(x)$ versus $x$, for $S=0$ (infinite square well or ISW with infinite walls at $x=0,a$) and $S=1,2,3,4$ (solid curves), along with low-lying
energy levels. The ground-state energy of the ISW, $E_{0}^{(S=0)} = {\cal E}_0$, is shown as the bottom horizontal dashed line and the ISW is the only
state for which that is a solution. Higher energy levels, such as the one labeled $E_2^{(S=0)} = E_1^{(S=1)} = E_0^{(S=2)} = 9{\cal E}_0$, appear as solutions for
more than one value of $S$.}
\label{fig:FIG1}
\end{figure}

\newpage

\begin{figure}[h!]
\epsfig{file=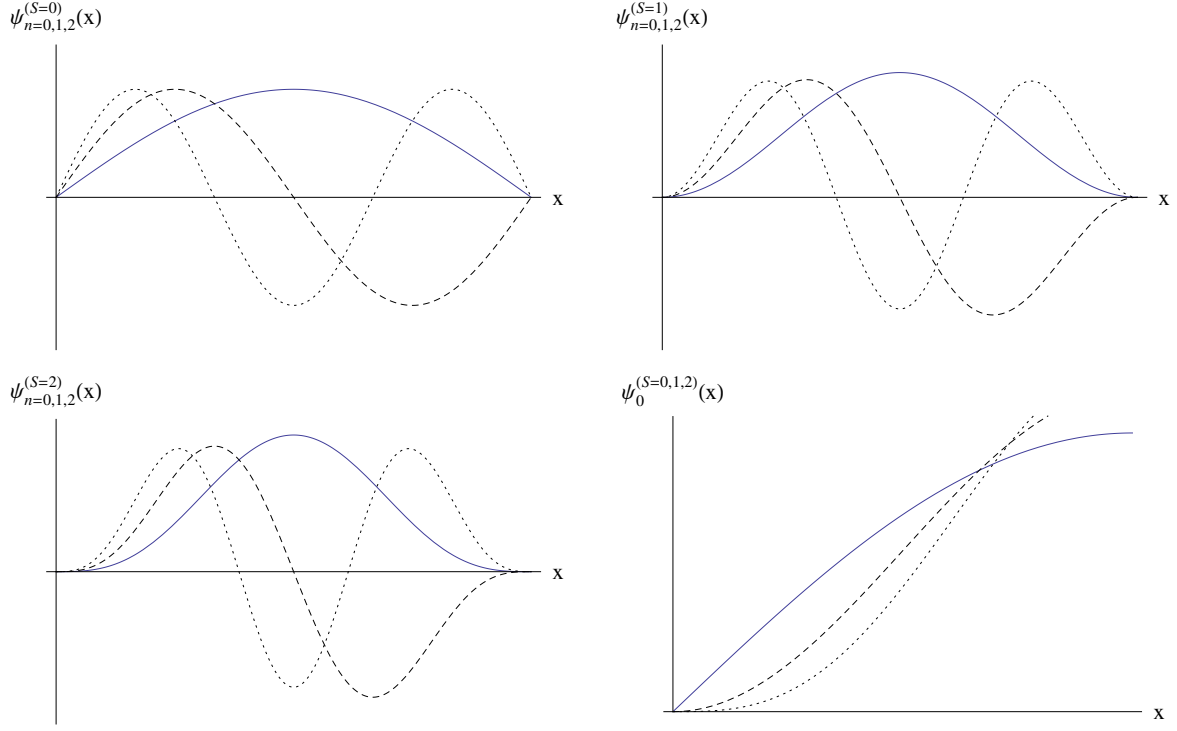,width=0.95\linewidth,angle=0}
\caption{Position-space solutions, $\psi_n^{(S)}(x)$ versus $x$, for $n=0,1,2$ (solid, dashed, dotted curves) for $S=0$ or ISW case (upper left),
$S=1$ (upper right), $S=2$ (lower left). In the lower right we show the ground-state solutions, $\psi_{0}^{(S)}(x)$, for $S=0,1,2$ (solid,
dashed, dotted curves) to illustrate the $x^{S+1}$ behavior near the infinite walls, as described near the end of Sec.~III.}
\label{fig:FIG2}
\end{figure}

\newpage

\begin{figure}[h!]
\epsfig{file=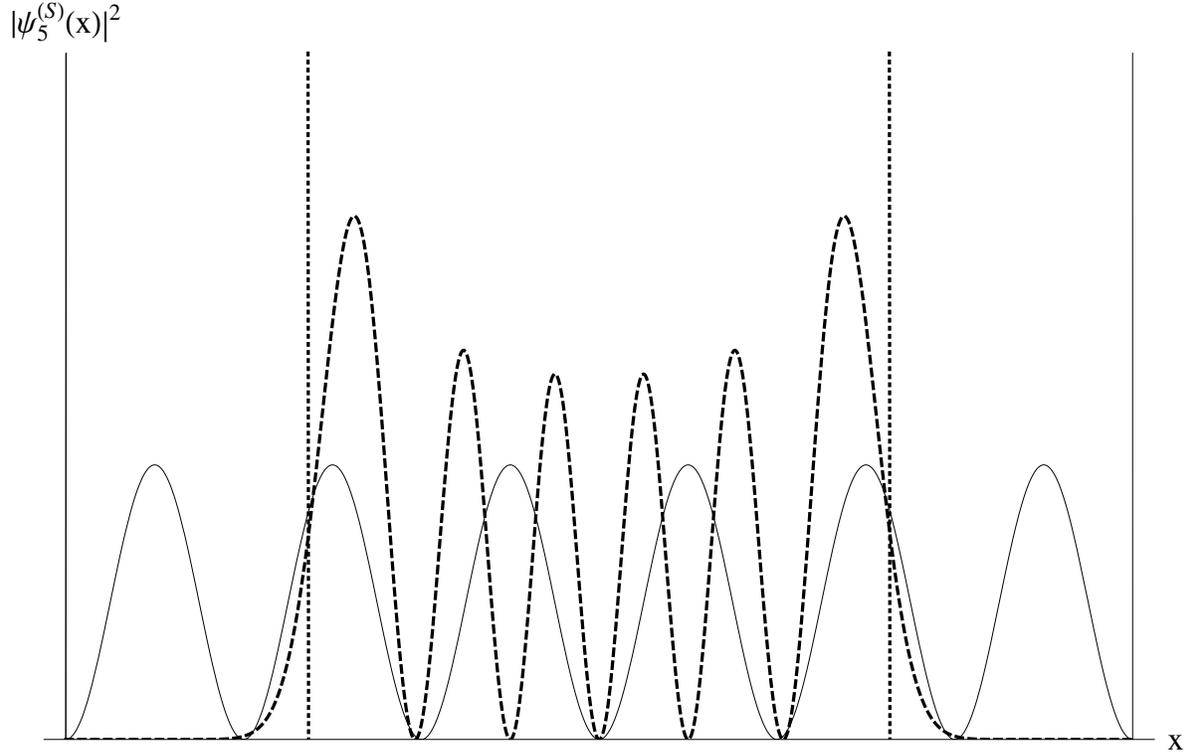,width=0.95\linewidth,angle=0}
\caption{Probability density, $|\psi_n^{(S)}(x)|^2$ versus $x$, for $n=5$, for $S=0,10$ (solid, dashed curves). For the $S=0$ or ISW case,
the probability density approaches the classical`'flat'  limit (after local averaging) for large $n$, while for $S>0$, the peaking of the probability density near the classical turning points of $V^{(S)}(x)$ is clear. For example, the bold vertical dotted lines indicate the classical turning points for the
$(n,S) = (5,10)$ case.}
\label{fig:FIG3}
\end{figure}

\end{document}